\documentclass[a4paper]{article}
\usepackage{graphicx}
\usepackage{subfiles}
\usepackage{amssymb}
\usepackage{textcomp}
\usepackage{amsmath}
\usepackage[ruled,vlined]{algorithm2e}
\begin{document}

\title{The Improvement of Decision Tree Construction Algorithm \\ Based On Quantum Heuristic Algorithms}

\author{
Ilnaz Mannapov \\ Kazan Federal University \\
                18,  Kremlyovskaya st, Kazan, Russia, 420008 \\ ilnaztatar5@gmail.com 
}
\date{}
\newtheorem{definition}{Definition}
\maketitle
\begin{abstract}
This work is related to the implementation of a decision tree construction algorithm on a quantum simulator. Here we consider an algorithm based on a binary criterion. Also, we study the improvement capability with quantum heuristic QAOA. We implemented the classical and the quantum version of this algorithm to compare built trees.

\end{abstract}

\section{Introduction}

Machine Learning is one of the famous directions of artificial intelligence \cite{mahesh2020machine}. The application ability of ML is very wide. The count of problems solved by machine learning algorithms arises day by day. The data size processed by them becomes larger. As a result machine learning algorithms require more computing resources. It means that if the algorithm works faster then it can process more data. Quantum computers \cite{nielsen2002quantum, a2017, aazksw2019part1} potentially may be useful for improvement of these algorithms \cite{wittek2014quantum, dw2001, quantumzoo}.

Let us describe some information about using the quantum computation for machine learning problems \cite{aazksw2019part2, wittek2014quantum}. The quantum computation is based on the quantum mechanics theory. The main notion of quantum computers ability is quantum parallelism. As result many instruments are invented based on this notion. Let us discuss about a few instruments. One of the famous technique used for speeding up the machine leatning algorithms is the Grover Search Algorithm \cite{g96,bbht98}. This technique give the quadratically speed up. More information about usage of Grover's algorithm can be found in the paper \cite{wittek2014quantum}. Another examples of instruments used in quantum machine learning are simulated quantum annealing \cite{wittek2014quantum}, efficient calculation of classical distances on a quantum computer \cite{qml}, SWAP-test \cite{aazksw2019part1} etc.

The goal of this work is to construct the improved version of the decision tree construction algorithm based on a binary criterion \cite{dectreesbook}. Also, we analyze the models built by classical and quantum versions of algorithms.

It should be noted that we do not consider the real quantum decision trees. For example, the paper \cite{lu2014quantum} proposes a quantum version of the decision tree. Their classifying process follows the classical algorithm with the only difference that we use quantum feature states encoding features into the states of a quantum system. At each node of the tree, the set of training quantum states is divided into subsets by a measurement \cite{qml, kms2021,ks2021,kms2019,ks2022}.

Our improvement works with classical decision trees. It is some subroutine that improves the construction of classical decision trees.

Nowadays decision trees \cite{quinlan} are not commonly used instruments for machine learning problems. However, wide-famous algorithms such as Random Forest and Gradient Tree Boosting are based on decision trees \cite{randomf}. As a result improvement of decision tree constructing algorithms could help solve more useful problems \cite{medforests}.

It should be noted that modern quantum computers are not applicable to show the exponential benefit of quantum algorithms. The technology of these devices is named as Noisy Intermediate-Scale Quantum (NISQ) technology \cite{preskill}. It is why we decided to consider algorithms that worked on such devices.

The QAOA is one of the famous algorithms implemented on NISQ devices. The QAOA is used for solving some optimization problems as MaxCut, MAX3-SAT. \cite{qaoa, qaoa2} In this paper, we consider this algorithm as an instrument of improving the decision tree constructing algorithm based on binary criteria. In the end, we show the experimental result of comparing the trees built by classical and quantum versions of the decision tree constructing algorithm.

The experiments show that the quantum algorithm builds the as same trees as a classical algorithm. Our previous work was related to the improvement of the decision tree constructing algorithms with impurity-based criteria. That work was based on the Grover's algorithm. In comparity of that paper in this investigation we shows that the quantum algorithms can help construct the same trees as the classical algorithms.

Section \ref{sec:prelims} is related to using decision trees for the classification problem. Also, we consider the different criteria of the decision tree construction algorithms. Section \ref{sec:idea} shows the idea used in our improvement. This section also provides some basics of QAOA. In Section \ref{sec:results}, we describe the comparative results of classical and quantum versions of the decision tree constructing algorithm based on the Twoing criterion.

\section{Preliminaries}\label{sec:prelims}

Machine learning allows us to predict a result using information about past events. The decision tree constructing algorithm is used to construct a decision tree for the classification problem. Let us formally consider a classification problem. 

There are two sequences: ${\cal X}=\{X^1, X^2, \dots, X^N\}$ is a training data set and ${\cal Y}=\{y_1, y_2, ..., y_N\}$ is a set of corresponding classes.  Here $X^i=\{x^i_1, x^i_2, ..., x^i_d\}$ is a vector of attributes, where $i\in\{1,\dots,N\}$, $d$ is a number of attributes, $A = (a_1, a_2, ..., a_d)$ is the set of all attributes, $N$ is a number of vectors in the training data set, $M$ is a number of classes. An attribute $x^i_j$ is a real-valued variable or a categorical variable.  Let $DOM_j=\mathbb{R}$ if $x^i_j$  is a real value; and $DOM_j=\{1,\dots,T_j\}$ if $x_j$ is a categorical attribute, i.e. $x^i_j\in\{1,\dots,T_j\}$ for some integer $T_j$. Let $y_i \in C=\{1,\dots,M\}$ be an index of class of $X^i$. $\sigma_{y=c_{i}} {\cal X}$ is a subset from training set which elements are related to class with number $i$, $c_{i} \in C$. The problem is to construct a function $F:DOM_1\times\ldots\times DOM_d\to C$ that is called classifier. The function classifies a new vector $X=(x_1,\dots,x_d) \notin {\cal X}$. Let $CV$, $RV$ be the notations to define the set of the categorical and real-valued attributes respectively.

The goal of decision tree constructing algorithms is to find the optimal decision tree by minimizing the generalization error.  Several authors have shown that finding a minimal decision tree consistent with the training set is NP-hard \cite{dectreesbook}. The main procedure of this algorithm is ``growing``. 

Suppose $B$ is some test with outcomes ${b_1, b_2,\dots, b_t}$ in a node. Then, there are $t$ outgoing edges for the node. Each leaf is associated with a result class from $C$. The testing process is the following. We start test conditions from the root node and go by edges according to the result of the condition. The label on the reached leaf is the result.

Our algorithm uses some quantum algorithms as a subroutine, and the rest part is classical.

\begin{figure}[h!]
    \centering
    \includegraphics[width=200px]{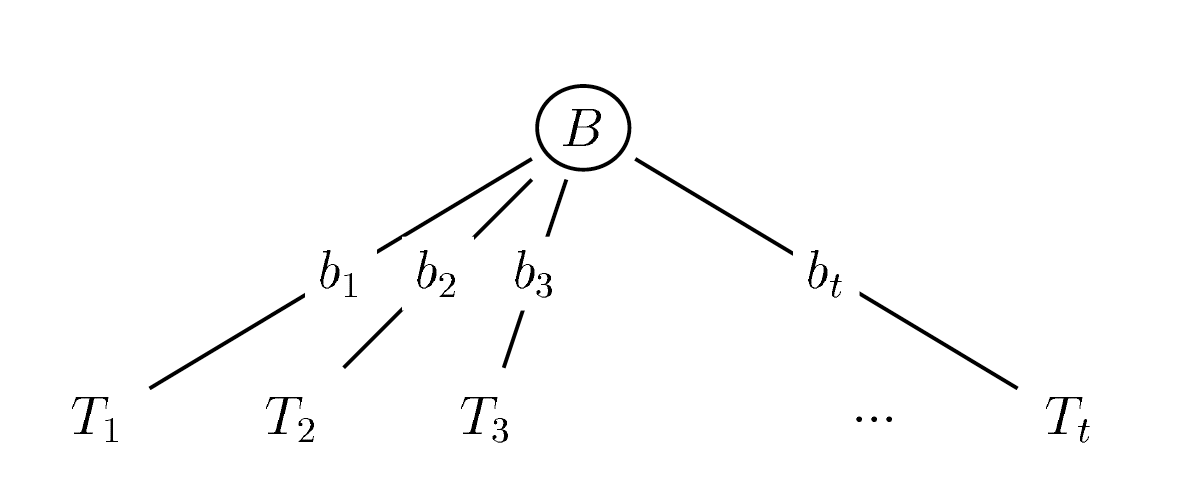}
    \caption{Testing $B$ \label{fig:dt}}
    \end{figure}

Decision tree learners use a method known as divide and conquer to construct a suitable tree from a training set ${\cal X}$ of vectors. If all vectors in ${\cal X}$ belong to the same class $c\in C$, then the decision tree is a leaf labeled by $c$. Otherwise, let $B$  be some test  (with outcomes ${b_1, b_2,\dots, b_t}$) that produces a non-trivial partition of ${\cal X}$. Let ${\cal X}_i$ be the set of training vectors from ${\cal X}$ that has outcome $b_i$ of $B$. Then, the tree is presented in Figure \ref{fig:dt}. Here $T_i$ is a result of growing a decision tree for a set ${\cal X}_i$.

The algorithm chooses some attribute with value to add a new node. This attribute value is used as a threshold for splitting data set into subsets. The process of selecting an attribute uses some split criteria. The algorithm maximizes its function.  

Many criteria are used for decision tree construction. The more popular and widely used criteria are defined in the next subsection.

\subsection{Criterion types}

\subsubsection{Impurity-based Criteria.}

Given a random variable $x$ with $k$ values, distributed according to $P=\left(p_{1}, p_{2}, \dots, p_{k}\right)$, an impurity measure is a function $\phi:[0,1]^{k} \rightarrow R$.

It should be noted that if the probability vector has a component of 1 (the variable $x$ gets only one value), then the variable is defined as pure. On the other hand, if all components are equal then the level of impurity reaches the maximum \cite{trees}.

\subsubsection{Normalized Impurity-based Criteria.}

The impurity-based criterion described above is biased towards attributes with larger domain values. Namely, it prefers input attributes with many values. Sometimes it is useful to "normalize" the impurity-based measures.

The famous decision tree constructing algorithms such as ID3, C4.5, C5.0, CART use impurity based-criteria, and normalized impurity-based criteria \cite{trees}. In the paper \cite{kms2021} we considered the classical and quantum improvement of the impurity-based decision tree construction algorithms.

\subsubsection{Binary Criteria.}

The binary criteria are used for creating binary decision trees. These measures are based on a division of the input attribute domain into two subdomains.

Let $\beta\left(a_{i}, d_{1}, d_{2}, S\right)$ denote the binary criterion value for attribute  $a_{i}$ over sample $S$ when $d_{1}$ and $d_{2}$ are its corresponded subdomains.

The value obtained for the optimal division of the attribute domain into two mutually exclusive and exhaustive subdomains, is used for comparing attributes, namely

\[
\begin{aligned}
&\beta^{*}\left(a_{i}, S\right)=\max \beta\left(a_{i}, d_{1}, d_{2}, S\right) \\
&\text { s.t. } \\
&d_{1} \cup d_{2}=\operatorname{DOM}\left(a_{i}\right) \\
&d_{1} \cap d_{2}=\emptyset
\end{aligned}
\]

In this work we consider one of the binary criteria named as Twoing \cite{trees}, \cite{breiman2017classification}. By $\left|\sigma_{C} S\right|$ the subdomain size of domain $S$ satisfied to condition $C$ is denoted.

\begin{equation}
\begin{aligned}
\text{twoing}\left(a_{i}, d_{1}, d_{2}, S\right) \\
& = 0.25 \cdot \frac{\left|\sigma_{a_{i} \in d_{1}} S\right|}{|S|} \cdot \frac{\left|\sigma_{a_{i} \in d_{2}} S\right|}{|S|} \\
& \cdot\left(\sum_{c_{j} \in \operatorname{DOM}(y)}\left|\frac{\mid \sigma_{a_{i} \in d_{1}} \text { and } y=c_{j} S \mid}{\left|\sigma_{a_{i} \in d_{1}} S\right|}-\frac{\mid \sigma_{a_{i} \in d_{2}} \text { and } y=c_{j} S \mid}{\left|\sigma_{a_{i} \in d_{2}} S\right|}\right|\right)^{2}
\label{eq1}
\end{aligned}
\end{equation}

\subsection{Classical algorithm}

On each step, the algorithm computes iteratively some values as subsets size distributed by classes, impurity function, etc.

Our algorithm deal with real-valued and discrete-valued (categorical) attributes. The pseudocode of the main procedure \textsc{TreeGrowing} is described by pseudocode (Algorithm \ref{alg:treegrowing}). 

\textsc{ChooseSplit} is the function that choose the split on each node (Algorithm \ref{alg:choose-split}).

The last but not least procedure is \textsc{SplitCriterion} (Algorithm \ref{alg:process-attribute}). 

All of these procedures are defined in the paper \cite{kms2021}.

\begin{algorithm}
\SetKwFunction{KwFn}{TreeGrowing}
\caption{Procedure of construction decision tree}
\label{alg:treegrowing}
\SetAlgoLined
\KwFn{$X'$}\\
\KwResult {Constructed decision tree for current node}
$T \leftarrow Tree()$ 
\tcp*[h]{a new tree with a root node is created}\;
\If {StopCriterion($X'$)}{
calculates amount of elements in ${\cal X}'$ by classes and finds the most common class $C_{cmn}$ in $O(|{\cal X}'|)$\;
$T.isLeaf \leftarrow true$\;
$T.label \leftarrow C_{cmn}$\;
}
\Else{
$max \leftarrow 0, attr \leftarrow \texttt{nil}, split \leftarrow \texttt{nil}, threshold \leftarrow -1$\;
$(attr, max, split, threshold) \leftarrow \textsc{ChooseSplit(X')}$\;
	\If {$attr \in CV$}{
		$split \gets$ \{split ${\cal X}'$ to subsets by attribute values\}\;
		$T.children \gets [\texttt{nil}, \dots, \texttt{nil}]$\;
		\For{$v_i \in DOM_{attr}$}{
			$ST \gets \textsc{TreeGrowing}(split[i])$\;
			$T.children[i] \leftarrow$ \{"$arg=v_i$", $ST$\}\;
		}
	}
	\Else{
		$T.children \leftarrow [\texttt{nil}, \texttt{nil}]$\;
         $ST_0 \gets \textsc{TreeGrowing}(split[0])$\;
		$ST_1 \gets \textsc{TreeGrowing}(split[1])$\;
		$T.children[0] \leftarrow$ ("$arg < threshold$", $ST_0$)\;
		$T.children[1] \leftarrow$ ("$arg \geq threshold$", $ST_1$)\;
	}
}
\Return $T$\;
\end{algorithm}

\begin{algorithm}
    \SetKwFunction{KwChoose}{ChooseSplit}
         \caption{The split choosing algorithm}
         \label{alg:choose-split}
         \KwChoose{$X'$}\\
         \KwResult {The best split}
           	$max \leftarrow 0, attr \leftarrow \texttt{nil}, split \leftarrow \texttt{nil}, threshold \leftarrow -1$\;
           	\For{$a \in A$}{
                $(cmax, csplit, cth) \leftarrow \textsc{SplitCriterion}(a, X')$\;
                \If{$cmax > max$}{
                    $max \leftarrow cmax, attr \leftarrow a, split \leftarrow csplit, threshold \leftarrow cth$\;
                }
            }
            \Return $(attr, cmax, csplit, cth)$
\end{algorithm}

\begin{algorithm}[H]
     \SetKwFunction{KwProcessing}{SplitCriterion}
         \caption{The attribute processing procedure}
         \label{alg:process-attribute}
         \KwProcessing{$attr, X'$}\\
         \KwResult {Data of processed attribute}
            \If{$attr \in CV$}
          {
                    $(cmax, csplit, cth) \gets \textsc{ProcessCategorical}(X', attr)$
             }
             \Else{
                    $(cmax, csplit, cth) \gets \textsc{ProcessReal}(X', attr)$
              }
        \Return $(cmax, csplit, cth)$
\end{algorithm}

Let us consider the last function (Algorithm \ref{alg:process-attribute}) in detail. The subroutine \textsc{SplitCriterion} processes an attribute differently depending on the type. We skip considering processing real-valued attributes because the quantum improvement is applied to the discrete-valued attributes subroutine.

Let us consider Equation \eqref{eq1}. As we can see adding a new node algorithm splits the training set into two subsets by maximizing Equation \eqref{eq1}. It should be noted that time complexity grows exponentially from the values count of the considered attribute.

\section{The Idea and The Implementation}\label{sec:idea}

We decide to use quantum heuristic QAOA to speed up calculating \eqref{eq1} for discrete-valued attributes.

\subsection{Quantum Approximate Optimization Algorithm}

Let us consider QAOA in details \cite{qaoa, qaoa2}. QAOA (Quantum Approximate Optimization Algorithm) is a quantum gate model algorithm to solve combinatorial optimization problems. The performance of the $p$-level QAOA ($QAOA_p$) increases continually along with $p$. Furthermore, one additional benefit of QAOA is fundamentally based on its simple structure. This simplicity leads to the use of QAOA on Noisy Intermediate-Scale Quantum (NISQ) devices \cite{preskill}.

\begin{definition}
In a combinatorial optimization problem defined on $n$-bit binary strings $z$, the objective function is defined as follows:

\begin{equation}
\label{eq3}
f(z):\{0,1\}^{n} \rightarrow R
\end{equation}

\end{definition}

\begin{definition}
We can map the objective function (\ref{eq3}) to the phase Hamiltonian, thus finding the optimal value of the objective function is a special case of finding the extremal eigenvalues for the phase Hamiltonian. The phase Hamiltonian $H_P$ encodes the objective function $f$ and acts diagonally on the computational basis states of $2_n$ dimensional Hilbert space ($n$-qubit space).

\[
H_{P}|z\rangle=f(z)|z\rangle
\]

In addition, the phase operators are defined as follows:

\[
U_{P}(\gamma)=e^{-i \gamma H_{P}}
\]

where $\gamma$ is a parameter.

\end{definition}

\begin{definition}

The mixing Hamiltonian $H_M$ is defined as follows:

\[
H_{M}=\sum_{j=1}^{n} \sigma_{j}^{x}
\]

where ${\sigma_j}^x$ is the Pauli-$X$ operator and $n$ is identical to the $n$ in (\ref{eq3}). In the quantum mechanical systems, the Pauli-$X$ operator acts as the $NOT$ operator, i.e., ${\sigma_j}^x|1\rangle = |0\rangle$ and ${\sigma_j}^x|0\rangle = |1\rangle$. In addition, the mixing operators are defined as follows:

\[
U_{M}(\beta)=e^{-i \beta H_{M}}
\]

where $\beta$ is a parameter.

\end{definition}

\begin{definition}
The initial state is as follows according to the superposition principle:

\[
|s\rangle=|+\rangle^{\otimes n}=\frac{1}{\sqrt{2^{n}}} \sum_{z}|z\rangle
\]

\end{definition}
Based on the definitions above, we can define the state of the $p$-level QAOA by applying the phase operator and the mixing operator alternately, as follows:

\begin{equation}
\label{eq9}
|\gamma, \beta\rangle=U_{M}\left(\beta_{p}\right) U_{P}\left(\gamma_{p}\right) \cdots U_{M}\left(\beta_{1}\right) U_{P}\left(\gamma_{1}\right)|s\rangle
\end{equation}

with an integer $p \geq 1$ and $2p$ parameters $\gamma_1 \cdots \gamma_p \equiv \gamma$ and $\beta_1 \cdots \beta_p \equiv \beta$. After the measurements in the computational basis are repeatedly performed in this state \ref{eq9}, the expectation value of $H_P$ can be obtained as follows:

\[
\left\langle H_{P}\right\rangle:=\left\langle\gamma, \beta\left|H_{P}\right| \gamma, \beta\right\rangle=\langle f\rangle_{(\gamma, \beta)}
\]

where $\langle f \rangle$ is the expectation value of the objective function (\ref{eq3}). The maximum or minimum value of $\langle H_P\rangle$ can be obtained by repeating the process of finding the optimal values of the parameters $\gamma$ and $\beta$. The iterative process for finding optimal parameters uses classical optimization methods. For this reason, QAOA is in the category of hybrid quantum-classical algorithms\cite{qaoa2}.

It should be noted that in our experiments the parameters $\gamma$ and $\beta$ were set empirically. The parameters defined one time before tree growing process. The QAOA implementation is called one time for each attribute processing.

\subsection{Implementation}

To check the quality of the algorithm we code it. The Kotlin programming language is used for this aim. We use several data set to compare the trees constructed by the classical and quantum version of the algorithm.

We implement a classical version of the decision tree construction algorithm which uses Twoing binary criteria as a split condition. Also, we code its quantum version which discrete-valued attributes processing is improved by QAOA.

We decide to implement our simulator. The reason is we need to use classical and quantum computation together. Unfortunately, the simulator cannot be used to estimate the speedup of the algorithm. The count of angles $p$ is set to $5$.

\section{Experiments}\label{sec:results}

Let us describe the experiments done for check the quality of quantum subroutine. As noted above, for this aim, we implemented the simulation framework of the classical and quantum versions of the decision tree construction algorithm. 

Let us describe some definitions of our comparing methodology. The ratio of the same nodes count to all nodes count is used to estimate the quality of improvement: 
\[
Q_{tree} = \frac{B_{eq}}{B},
\]

where $B$ is all nodes count, $B_{eq}$ is the same nodes count.

The finish of the tree construction process depends on the data and the height parameter. In our experiment we use the next height parameters: $h = 3$, $h = 5$, $h = 7$, $h = 10$, $h = 15$.

It should be noted that we could use more values of the height parameter.

The experiments were made for several datasets. As described above our improvement works for categorical attributes. Because of this, for check our result we should find datasets with categorical attributes. For this aim we looked the absolutely random data sets from the open sources. All of the datasets are available here \cite{datasets}.

\subsection{Experiment 1}

The first experiment is done for data set \cite{cars}. The size of this set is $1728$. The database contains examples with the structural information removed, i.e., directly relates CAR to the six input attributes: buying, maint, doors, persons, lug\_boot, safety. All attributes are categorical.

The values of $Q_{tree}$ are next:

\begin{tabular}{ | l | l| }
\hline
The height of tree $h$ & $Q_{tree}$  \\ \hline
$h = 3$ & $1.0$ \\\hline
$h = 5$ & $1.0$ \\\hline
$h = 7$ & $1.0$ \\\hline
\end{tabular}

\subsection{Experiment 2}

The second experiment is done for data set \cite{nursery}. Nursery Database was derived from a hierarchical decision model originally developed to rank applications for nursery schools. It was used during several years in 1980's when there was excessive enrollment to these schools in Ljubljana, Slovenia, and the rejected applications frequently needed an objective explanation. The final decision depended on three subproblems: occupation of parents and child's nursery, family structure and financial standing, and social and health picture of the family. The size of this set is $12960$. An attributes count is $8$. All attributes are categorical.

The values of $Q_{tree}$ are next:

\begin{tabular}{ | l | l| }
\hline
The height of tree $h$ & $Q_{tree}$  \\ \hline
$h = 3$ & $1.0$ \\\hline
$h = 5$ & $1.0$ \\\hline
$h = 7$ & $1.0$ \\\hline
$h = 10$ & $1.0$  \\\hline
\end{tabular}

\subsection{Experiment 3}

The third experiment is done for data set \cite{abalone}. Predicting the age of abalone from physical measurements. The age of abalone is determined by cutting the shell through the cone, staining it, and counting the number of rings through a microscope -- a boring and time-consuming task. The size of this set is $4177$. An attributes count is $8$. It should be noted that this dataset contains mixed attributes as categorical and real-valued.

The values of $Q_{tree}$ are next:

\begin{tabular}{ | l | l| }
\hline
The height of tree $h$ & $Q_{tree}$  \\ \hline
$h = 3$ & $1.0$ \\\hline
$h = 5$ & $1.0$ \\\hline
$h = 7$ & $1.0$ \\\hline
$h = 10$ & $1.0$  \\\hline
\end{tabular}

\subsection{Experiment 4}

This experiment is done for data set \cite{cnt}. This database contains all legal 8-ply positions in the game of connect-4 in which neither player has won yet, and in which the next move is not forced.
The size of this set is $67557$. An attributes count is $42$. All attributes are categorical.

The values of $Q_{tree}$ are next:

\begin{tabular}{ | l | l| }
\hline
The height of tree $h$ & $Q_{tree}$  \\ \hline
$h = 3$ & $1.0$ \\\hline
$h = 5$ & $1.0$ \\\hline
$h = 7$ & $1.0$ \\\hline
$h = 10$ & $1.0$  \\\hline
$h = 15$ & $1.0$  \\\hline
\end{tabular}

\subsection{Experiment results}

The results demonstrate that the constructed trees are equal for the classical and quantum versions of the algorithm. The code of our experiments is provided here \cite{gitlab}. The code was written in Kotlin.

The structure of code is next. The file 'Data.kt', 'ComplexNumber.kt' contains classes used for the tree constructing. Also in 'Data.kt' the input data parsing method is provided. The classical and quantum algorithms implementation is described in 'Tree.kt' and 'Quantum.kt' respectively.

The experiments was made on the laptop with processor Intel Core I5.

\section{Conclusion}

We considered an algorithm that uses the quantum subroutine for maximization impurity value for discrete-valued attributes. The experiment shows that the tree constructed by the quantum algorithm is identical to a tree by the classical version. The speedup of quantum can be proven theoretically. These results confirm the practical usability of quantum computers for ML problems. We have some open questions. 
\begin{itemize}
    \item How to compare the running time of classical and quantum implementation of algorithms?
    \item Which heuristic should be used to set optimal QAOA parameters?
\end{itemize}

\paragraph*{Acknowledgements.}
Kazan Federal University for the state assignment in the sphere of
scientific activities, project No. 0671-2020-0065.

%
%

\bibliographystyle{plain} 

\end{document}